\documentclass[journal]{IEEEtran}
\usepackage{graphicx}
\usepackage{epsf}
\usepackage{mathtools}
\usepackage{amssymb}
\usepackage{amsmath, amsfonts}
\usepackage{latexsym}
\usepackage{subfigure}
\usepackage{multirow}
\usepackage{multicol}
\usepackage[table]{xcolor}
\usepackage{color}

\usepackage{tabularx}
\usepackage{lipsum}

\usepackage{algorithm}
\usepackage{algorithmic}

\usepackage{mathrsfs}

\usepackage{fixltx2e}

\usepackage[mathscr]{euscript}

\usepackage{commath}
\usepackage{MnSymbol}

\usepackage{subfigure}

\usepackage{mathtools} 
\DeclarePairedDelimiterX\setc[2]{[}{]}{\,#1 \;\delimsize\vert\; #2\,}
\DeclarePairedDelimiterX\parth[2]{(}{)}{\,#1 \;\delimsize\vert\; #2\,}

\PassOptionsToPackage{hyphens}{url}
\usepackage[unicode=true, bookmarks=true, bookmarksnumbered=true, bookmarksopen=true, bookmarksopenlevel=1, breaklinks=false, pdfborder={0 0 0}, pdfborderstyle={}, backref=false, colorlinks=false]{hyperref}

\usepackage{braket}

\definecolor{orange}{RGB}{255,127,0}
\definecolor{blue}{RGB}{0,0,255}
\definecolor{red}{RGB}{255,0,0}
\definecolor{green}{RGB}{50,160,50}
\definecolor{grey}{RGB}{125,120,125}
\definecolor{purple}{RGB}{125,0,125}

\begin{document}
{
\title{\fontsize{24}{28}\selectfont Adding Interactivity to Education of\\Complex Wireless Networks\\Using Digital Game-Based Learning}

\author
{
\vspace{0.2 in}
Seungmo Kim, \textit{Senior Member}, \textit{IEEE}

\thanks{S. Kim is with the Department of Electrical and Computer Engineering, Georgia Southern University in Statesboro, GA, USA. The corresponding author is S. Kim who can be reached at seungmokim@georgiasouthern.edu.}
}

\maketitle
\begin{abstract}
Can we make undergraduate engineering education easier and more fun? This research aims to see if we can answer the ambitious question! The digital game-based learning (DGBL) has been found to increase the efficacy of learning when applied in engineering classes thanks to its ability to make students feel easy and fun. However, the state-of-the-art DGBL schemes still observe challenges in various aspects including cost, efficacy, readiness of instructors and students, etc. Motivated from the challenges, this research proposes to design a DGBL platform that features visualized and systematic views. Specifically, we identify the blockchain applied to wireless communications and networking systems as a key ecosystem that we capitalize the benefit of the proposed platform.. As such, in this paper, we lay out a comprehensive DGBL pedagogy that includes (i) creation of relevant assignment activities and class materials in a relevant course and (ii) evaluation of the pedagogical efficacy. In a long-term view, a successful delivery of this research will increase the confidence of undergraduate engineering students on the ``in-concert'' dynamics of various factors determining the performance of a blockchain system built on a wireless network.
\end{abstract}


\section{Introduction}\label{sec_intro}
This paper certifies request of presentation at the `Innovative in Practice' track of the `Work in Progress (WIP)' session.

\subsection{Motivation}\label{sec_intro_motivation}
There is a jumble of symbols, concepts, systems, and protocols that make engineering and science sound very vague and incomprehensive. The question we raise is: \textit{Can we provide effective access and understanding of these concepts and introduce them into an undergraduate curriculum?}

Digital game is one of the most efficient tools to motivate youngsters: they feel easy, keen, confident, and fun in dealing with games, which makes them enjoy thriving to go through challenges. Taking advantage of this, the digital game-based learning (DGBL) is known to make learning more attractive, motivating, and personalized from the learner's viewpoint \cite{iuse_1}\cite{iuse_2}. Some difficulties in carrying out typical forms of inquiry-based learning in a class have been pointed out \cite{iuse_3}-\cite{iuse_6}: to summarize, (i) the complexity in the contexts and (ii) the students' low engagement level. However, when delivered in the form of a digital game-based activity, greater chances of being more joyful and efficient have been found: specifically, (i) self-discovery and construction of one's own reality \cite{iuse_7}-\cite{iuse_11}; (ii) increase in students' inquiry experience \cite{iuse_12}\cite{iuse_13}; (iii) engagement of students in applying knowledge to real-world contexts \cite{iuse_14}-\cite{iuse_19}; and (iv) collaborative learning where two or more learners interact and engage to learn together \cite{iuse_20}-\cite{iuse_22}.

In further studies, it has been found that the level of students' engagement influences their inquiry learning effectiveness \cite{iuse_3}\cite{iuse_23}-\cite{iuse_25}. It motivates the need for precise evaluation of students' feedback on management of the proposed gamified learning platform (as shall be described in Section \ref{sec_conclusions}).

We identify \textit{blockchain applied in wireless communications} as the new content to explore the effectiveness of DGBL in an undergraduate curriculum. The key rationale is that this rapidly evolving field of technology offers great opportunities for students to experience multidisciplinary topics involving elements of mathematics, (i.e., abstract algebra and number theory), cryptography, computer networks, etc. The necessity for \textit{systematic} comprehension of such an interplay among various, dissimilar areas makes a compelling case in favor of infusing new learning materials and strategies to enhance an undergraduate education program. In fact, the wireless technology is experiencing a variety of completely new problem domains including spectrum sharing \cite{lett17} and vehicle-to-everything (V2X) networking \cite{arxiv19}\cite{dave}. Nonetheless, at the blockchain consensus 2017, scholars have concluded that \textit{academics are not keeping pace with blockchain change} \cite{iuse_27}.

Given the significance of blockchain expertise in new undergraduates, strong educational support is vital and current educational curricula should reflect cutting-edge trends and needs in this sector. Specifically, students are challenged more than ever to be creative to confront contemporary issues related to blockchain. Hence, educational environments should cultivate students that are equipped with a set of tools to formulate, solve, and properly tackle multidisciplinary problems.

\subsection{Contributions}\label{sec_intro_contributions}
While playing on the proposed gamified platform, students are expected to learn on such multifaceted interplay that is enabled by \textit{visualization} of cause and effect of different approaches and decisions that they make. For instance, the \textit{connectivity} is a critical factor determining the performance of a network, which is not a straightforward concept to grasp when not visualized. In the proposed platform, through visualization, students will be able to (i) perceive which areas are better or less connected, (ii) route their own nodes toward the directions that they remain in a maximum connectivity, and (iii) in turn, maximize the data exchange performance through a round of the game. Such pedagogical improvement led by this research will provide a model for similar approaches in other challenging science, technology, engineering, and mathematics (STEM) education domains.

\section{State-of-the-Art at Other Institutions and Their Limitations}\label{sec_related}
\subsection{Current Undergraduate Programs and Limitations}
Several institutions have deployed DGBL in their undergraduate curricula. We discuss their contributions and limitations as follows.

Virginia Tech \cite{iuse_28} has developed an online learning platform providing visualization and gamification and published advantages on education of wireless communications engineering \cite{iuse_45}. Yet, the need for connecting 48 physical hardware nodes, the system requires significant amounts of time and cost to maintain or host demonstrations or competitions. Moreover, it also left pedagogical challenges including: inabilities to visualize and analyze (i) \textit{mobility} of nodes in a network and (ii) \textit{systematic} perspectives as a result of interplay among multiple mobile nodes and parameters determining a network characteristic.

The Game Studio at American University \cite{iuse_29} offers academic programs (including baccalaureate and graduate) across interdisciplinary fields including human-computer interaction, civic engagement, journalism, art, education, and health. However, the published games' contents are not suitable for immediate adaptation to STEM education. Moreover, access to the actual games are not available for general users, which limits the applicability.

The City University of New York (CUNY) Games Network \cite{iuse_30} connects educators to games and other forms of interactive and inquiry-based learning. However, the games available on the network do not provide capabilities to understand \textit{systematic} interplay among multiple variables and factors, which is easily encountered in a STEM-related problem solving.

The Embry-Riddle Aeronautical University \cite{iuse_31} investigated efficacy of games in understanding human behavior--i.e., personality, motivation, performance, teamwork, etc. But, being dedicated to psychological disciplines only, their pedagogical findings have little implication to improvement of a STEM curriculum.

In Fall 2013, via the A-Games Project \cite{iuse_32}, the University of Michigan surveyed 488 K-12 teachers nationwide about the efficacy of DGBL. From uses of games to cover content mandated by state/national or local/district standards, the majority of teachers were found to believe games were effective for motivating students, helping students reinforce or master previously taught content, providing useful information about student learning, and teaching students new content.

\subsection{Common Challenges}
Some scholarly work has also studied general challenges found from other DGBL programs: (i) cost of games \cite{iuse_32}; (ii) limited time to deploy the games and analyze the efficacy \cite{iuse_32}; (iii) lack of technology literacy by teachers and students \cite{iuse_32}; (iv) lack of place to find appropriate games \cite{iuse_32}; (v) performance evaluation methods still relying on standardized test scores \cite{iuse_32}; (vi) necessity of additional resources (i.e., pedagogical and technical supports) \cite{iuse_33}; (vii) issues of access and the digital divide \cite{iuse_33}; (viii) necessity of periodic examination on whether students actually prefer this approach to teaching \cite{iuse_33}; and (iv) fear of distraction due to perception of a game as leisure rather than academic drive \cite{iuse_34}\cite{iuse_35}.

\section{Problem Statements and Respective Action Plans}\label{sec_problem}
The goal of this research is to modify current undergraduate electrical and computer engineering (ECE) education pedagogy to employ an innovative gamified tool, which will make concepts related to blockchain more accessible to undergraduates. While the DGBL has been found to be effective in education of many engineering aspects, this research is particularly focused on enhancing the efficacy of learning dynamic behaviors of complex wireless networks (e.g., V2X networks). The key rationale is the need for \textit{systematic} understanding of orchestration of various topics forming the basis of such a complex wireless network: e.g., probability and statistics, optimization, communications theory, networking theory, graph theory, cryptography, etc.

As a result of careful investigation about the currently available DGBL-related literature and resources, this research specifies three problem statements:
\begin{itemize}
\item \textbf{\textsf{P1:}} No DGBL focuses on training a \textit{systematic} view on complex behaviors of a network
\item \textbf{\textsf{P2:}} No explicit finding exists on the \textit{efficacy} of DGBL focusing on such systematic understanding
\item \textbf{\textsf{P3:}} It is critical to monitor \textit{students' perception} on a new learning tool
\end{itemize}
In response to the problem statements, we design this research in such way to undertake three action plans :
\begin{itemize}
\item \textbf{\textsf{A1:}} \textit{Development} of a gamified learning platform
\item \textbf{\textsf{A2:}} \textit{Incorporation} into a number of undergraduate courses
\item \textbf{\textsf{A3:}} \textit{Evaluation} of the efficacy of the proposed approach
\end{itemize}
As such, this research aims to transform a difficult-to-grasp field into an easily accessible learning environment, which will contribute to a breakthrough in ECE education.

\begin{figure}[hbtp]
\centering
\includegraphics[width = \linewidth]{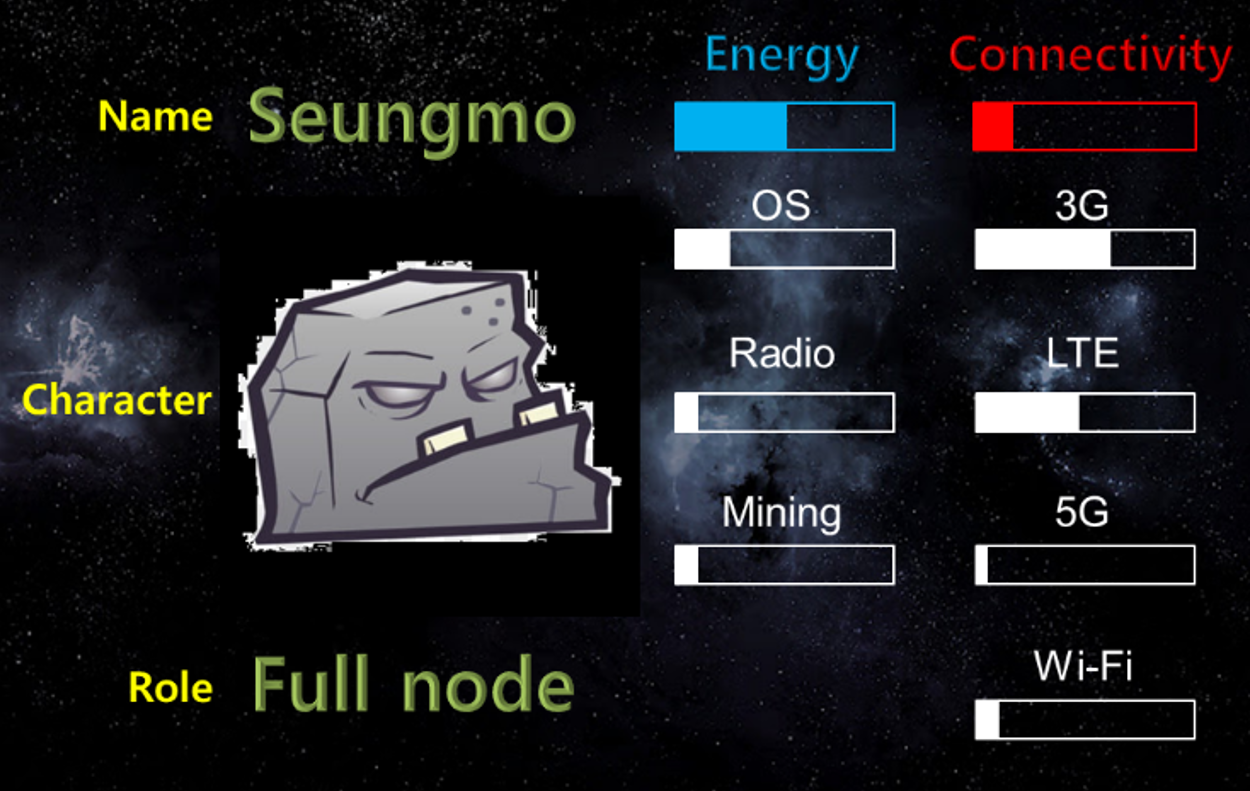}
\caption{Example of a character view}
\label{fig_seungmo}
\end{figure}

\begin{figure}[hbtp]
\centering
\includegraphics[width = \linewidth]{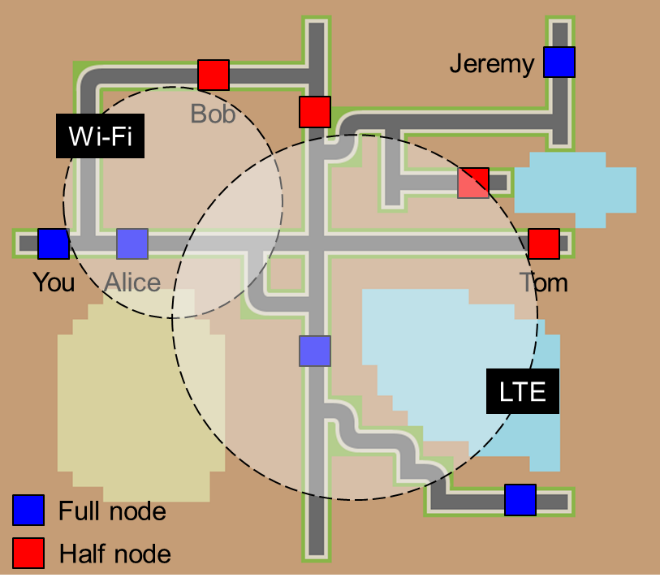}
\caption{Screenshot of an example map with nodes distributed}
\label{fig_map_nodes}
\end{figure}

\begin{figure}[hbtp] 
\centering
\includegraphics[width = \linewidth]{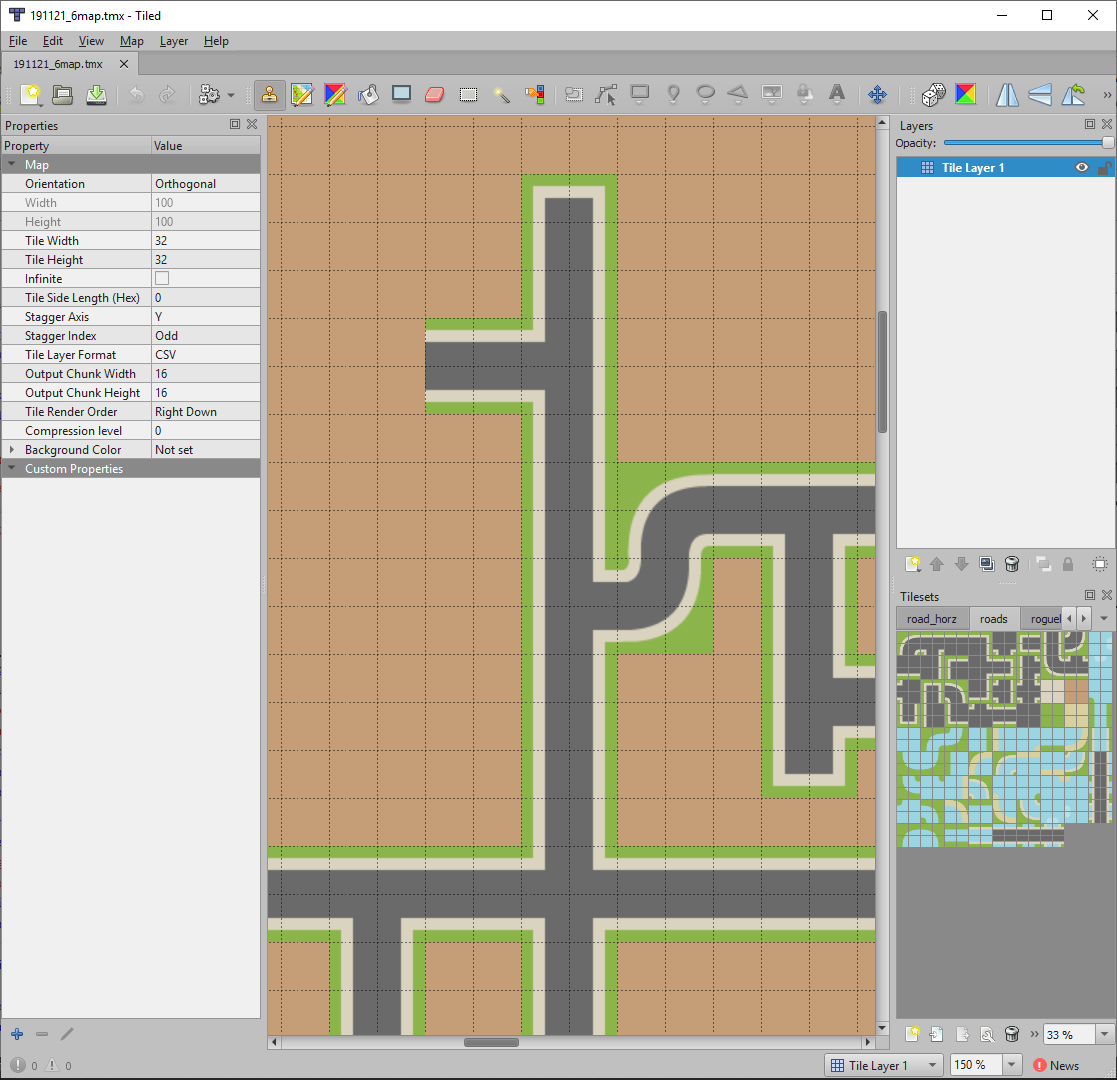}
\caption{Screenshot of platform development}
\label{fig_map_development}
\end{figure}

\begin{figure}[hbtp]
\centering
\includegraphics[width = \linewidth]{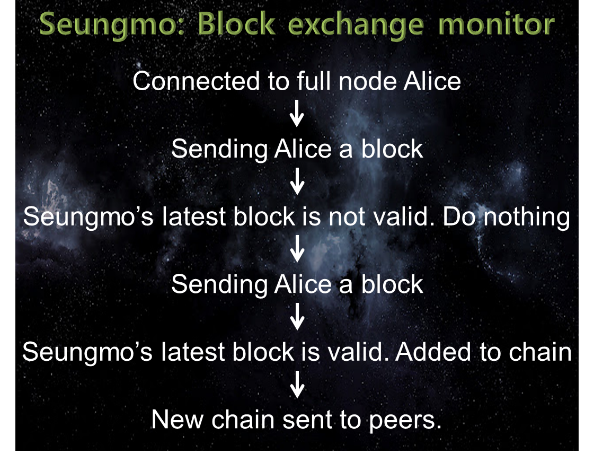}
\caption{Example of block management process}
\label{fig_process}
\end{figure}

\section{Current Status}\label{sec_status}
We report that the current status of this project is approximately 50\% toward the completion of \textbf{\textsf{A1}}. Technical details follow in this section.

\subsubsection{Player Character}\label{sec_status_development_character}
As illustrated in Fig. \ref{fig_seungmo}, each user is able to choose a character. Each character is assigned a certain role between a ``full'' node and a ``half'' node. A full node has full capabilities including block generation, mining (i.e., validation), and connection to networks. A half node, on the other hand, has all the capabilities but is not able to mine a block. Also, other capabilities (i.e., block mining speed, mobility speed, indoor-outdoor penetration, jump, etc) are differentiated among the characters.

While playing a game, a user sees a map to move its character, as shown in Fig. \ref{fig_map_nodes}. The user is able to switch the display into the “character view” as shown in Fig. \ref{fig_seungmo}, to check the current status of (i) energy consumption and (ii) connectivity. Note that energy consumption is monitored because it is one of the key parameters measuring the efficiency of an entire blockchain system \cite{iuse_55}, which will be explained in more depth in the Scenarios part of this section.

\subsubsection{Map}
The game will be designed as a role-playing game (RPG) \cite{iuse_56}, to focus on visualizing the interplay among parameters of network connectivity. As shown in Fig. \ref{fig_map_development}, we have already (i) started developing a game environment and thus (ii) is familiar with related tools to enhance the quality. A user is able to switch to the blockchain status as shown in Fig. \ref{fig_process}, which displays all the activities of (i) block generation, (ii) attempt of validation, (iii) result of validation, and (iv) addition to the chain.

\subsubsection{Connectivity}
For modeling signals' propagation, we adopt the standard models defined by the 3rd Generation Partnership Project (3GPP) \cite{iuse_54}, which determine the size of a cell and types of obstacle (e.g., cars, buildings, etc). While the geographical type can be selected by a user, as mentioned earlier in the Platform part, other geographical settings will be randomized in every round of the game--e.g., locations of roads, base stations, obstacles, etc. Also, the network type will be differentiated depending on the game character that a user has selected. For instance, character Alice can be connected to Wi-Fi, 3G, and 5G, while character Bob can be connected to Wi-Fi and Bluetooth. Lastly, network coverage will also be varied. Not all the area is being covered by the network coverages: some areas will be left to be uncovered, in which no block can be exchanged \cite{verboom}\cite{kabir}. Each user can make a strategy to route to maximize the coverage, which will in turn maximize the performance of a blockchain system.

\subsubsection{Scenarios}
Each student's character control the character on the \textit{roads} in the map as shown in Fig. \ref{fig_map_nodes}. A block is generated every $N$ seconds at a random user (no matter a full or a half node), and has to be ``validated'' to go through a PoW and be added to the chain \cite{iuse_57}. A block can only be validated by a full node \cite{iuse_58}, and thus must be handed to another full node for validation \cite{iuse_59}. The time until a rendezvous with a full node varies according to various parameters--e.g., connectivity, a character's moving speed, location on the road, number of obstacles, etc. The rationale is that it usually takes a tradeoff between the blockchain's pursuit of the distributed decision making and the wireless' complexity in dealing with the distributed networking architecture \cite{access19}.

\section{Pilot Plan}\label{sec_plan}
The outcome of this project will be \textit{incorporated} into interactive assignments and class materials and \textit{piloted} in a course that the author is currently teaching, in such a way to accomplish \textbf{\textsf{A2}}. We will develop two types of activities based on the proposed gamified learning platform--namely, \textit{interactive lab} session and \textit{hands-on assignment}. The following materials will be developed and used commonly in both types of the assignment: lecture, student report format, quiz, survey, and activity instruction manual.

The next step is to \textit{evaluate} the efficacy of the proposed pedagogy, which will accomplish \textbf{\textsf{A3}}. The impact of using the proposed interactive tool on student learning will be assessed based on the criteria of student learning outcomes (SLOs) for engineering by the Accreditation Board for Engineering and Technology (ABET) \cite{iuse_50}. The Computing Accreditation Commission (CAC) SLOs \cite{iuse_61} will be used as a secondary reference. Moreover, since the experiment will collect data from human participants, all experimental protocols will be reviewed and approved by both institutions' Institutional Review Boards (IRBs). The IRBs will confirm that the protocols are compliant with all state and federal regulations involving human subjects testing, ensuring proper care throughout the recruitment of subjects and collection of data.

\section{Conclusions}\label{sec_conclusions}
This research aims to (i) build a DGBL platform, (ii) incorporate into an ECE course, and (iii) evaluate the pedagogical efficacy. The project had been motivated from three problem statements, which corresponded to three action plans. As reported in this paper, we have made the progress through \textbf{\textsf{A1}}, while \textbf{\textsf{A2}} and \textbf{\textsf{A3}} are left clearly planned. To elaborate our progress in \textbf{\textsf{A1}}, we have built a DGBL framework that is similar to a RPG where multiple players can explore different maps with different levels of wireless connectivity. By doing so, students are expected to learn how the connectivity affects the performance of a wireless network and a blockchain system overlaid on the network. In conclusion, the success of this research will cultivate students that are equipped with skillsets to deal with a wide variety of problems raised in the rapidly evolving ecosystems of blockchain and wireless communications.



\begin{thebibliography}{99}
\setlength{\parskip}{0.0000001 em}

\bibitem{iuse_1} D. Kay and J. Kibble, ``Learning theories 101: application to everyday teaching and scholarship,'' \textit{Advanced Physiology Education}, vol. 40, 2016.

\bibitem{iuse_2} M. Prensky, ``Digital game-based learning,'' \textit{Computers in Entertainment (CIE)}, vol. 1, no. 1, Oct. 2003.

\bibitem{iuse_3} B. R. Lim, ``Challenges and issues in designing inquiry on the web,'' \textit{British Journal of Educational Technology}, vol. 35, no. 5, 2004.

\bibitem{iuse_4} M. Lowenstein, ``General education, advising, and integrative learning,'' \textit{The Journal of General Education}, vol. 64, no. 2, 2015.

\bibitem{iuse_5} C. A. Thompson, M. Eodice, and P. Tran, ``Student perceptions of general education requirements at a large public university: no surprises?'' \textit{The Journal of General Education}, vol. 64, no. 4, 2015.

\bibitem{iuse_6} S. Ucar and K. C. Trundle, ``Conducting guided inquiry in science classes using authentic, archived, web-based data,'' \textit{Computers \& Education}, vol. 57, no. 2, 2011.

\bibitem{iuse_7} M. Birenbaum, K. Breuer, E. Cascallar, F. Dochy, Y. Dori, and J. Ridgway, ``A learning integrated assessment system,'' \textit{Educational Research Review}, vol. 1, 2006.

\bibitem{iuse_8} M. Hannafin, K. Hannafin, S. Land, and K. Oliver, ``Grounded practice and the design of constructivist learning environments,'' \textit{Education Technology Research and Development}, vol. 45, no. 3, 1997.

\bibitem{iuse_9} M. T. Huber , C. Brown, P. Hutchings, R. Gale, R. Miller, and B. Breen, ``Integrative learning opportunities to connect,'' \textit{Association of American Colleges and Universities and The Carnegie Foundation for Advancement of Teaching 2007}.

\bibitem{iuse_10} D. Walker, ``Integrative education,'' \textit{Research Roundup}, vol. 12, no. 1, 1995.

\bibitem{iuse_11} J. R. Wingert, S. A. Wasileski, K. Peterson, L. G. Mathews, A. J. Lanou, and D. Clarke, ``Enhancing integrative experiences: Evidence of student perceptions of learning gains from cross-course interactions,'' \textit{Journal of Scholarship of Teaching and Learning}, vol. 11, no. 3, 2011.

\bibitem{iuse_12} F. Giannakas, G. Kambourakis, and S. Gritzalis, ``CyberAware: A mobile game-based app for cybersecurity education and awareness,'' in \textit{Proc. IEEE International Conference on Interactive Mobile Communication Technologies and Learning (IMCL)}, Nov. 2015.

\bibitem{iuse_13} A. Proske, R. D. Roscoe, and D. S. McNamara, ``Game-based practice versus traditional practice in computer-based writing strategy training: Effects on motivation and achievement,'' \textit{Educational Technology Research and Development}, vol. 62, no. 5, 2014.

\bibitem{iuse_14} H. C. Chu, G. J. Hwang, C. C., Tsai, and J. C. R. Tseng, ``A two-tier test approach to developing location-aware mobile learning system for natural science course,'' \textit{Computers \& Education}, vol. 55, no. 4, 2010.

\bibitem{iuse_15} S. Erhel and E. Jamet, ``Digital game-based learning: Impact of instructions and feedback on motivation and learning effectiveness,'' \textit{Computers \& Education}, vol. 67, 2013.

\bibitem{iuse_16} H. Peng, Y. J. Su, C. Chou, and C. C. Tsai, ``Ubiquitous knowledge construction: Mobile learning re-defined and a conceptual framework,'' \textit{Innovations in Education and Teaching International}, vol. 46, no. 2, 2009.

\bibitem{iuse_17} M. Ruchter, B. Klar, and W. Geiger, ``Comparing the effects of mobile computers and traditional approaches in environmental education,'' \textit{Computers \& Education}, vol. 54, no. 4, 2010.

\bibitem{iuse_18} J. Sanchez and R. Olivares, ``Problem solving and collaboration using mobile serious games,'' \textit{Computers \& Education}, vol. 57, no. 3, 2011.

\bibitem{iuse_19} B. Vogel, A. Kurti, M. Milrad, E. Johansson, M. Muller, ``Mobile inquiry learning in Sweden: Development insights on interoperability, extensibility and sustainability of the LETS GO software system,'' \textit{Journal of Educational Technology \& Society}, vol. 17, no. 2, 2014.

\bibitem{iuse_20} S. Amara, J. Macedo, F. Bendella, and A. Santos, ``Group formation in mobile computer supported collaborative learning contexts: A systematic literature review,'' \textit{Education Technology \& Society}, vol. 19, no. 2, 2016.

\bibitem{iuse_21} S. T. Cavusgil, R. J. Calantone, and Y. Zhao, ``Tacit knowledge transfer and firm innovation capability,'' \textit{Journal of Business and Industrial Marketing}, vol. 18, no. 1, 2003.

\bibitem{iuse_22} P. Dillenbourg, ``What do you mean by collaborative learning?'' \textit{Collaborative Learning: Cognitive and Computational Approaches}, Oxford: Elsevier, 1999. [Online]. Available: \url{https://telearn.archives-ouvertes.fr/hal-00190240/document}

\bibitem{iuse_23} A. All, E. P. Nunez Castellar, and J. Van Looy, ``Assessing the effectiveness of digital game-based learning: Best practices,'' \textit{Elsevier Computers \& Education}, vol. 92, Jan. 2016.

\bibitem{iuse_24} G.-J. Hwang and C.-H. Chen, ``Influences of an inquiry-based ubiquitous gaming design on students' learning achievements, motivation, behavioral patterns, and tendency towards critical thinking and problem solving,'' \textit{British Journal of Educational Technology}, vol. 48, no. 4, 2017.

\bibitem{iuse_25} N. Rutten, J. T. van der Veen, and W. R. van Joolingen, ``Inquiry-based whole-class teaching with computer simulations in physics,'' \textit{International Journal of Science Education}, vol. 37, no. 8, 2015.

\bibitem{lett17} S. Kim and C. Dietrich, ``Coexistence of outdoor Wi-Fi and radar at 3.5 GHz,'' \textit{IEEE Wireless Commun. Lett.}, vol. 6, iss. 4, Aug. 2017.

\bibitem{arxiv19} S. Kim and M. Bennis, ``Spatiotemporal analysis on broadcast performance of DSRC with external interference in 5.9 GHz band,'' \textit{arXiv:1912.02537}, Dec. 2019.

\bibitem{dave} T. Dessalgn and S. Kim, ``Danger aware vehicular networking,'' \textit{arxiv preprint arXiv:2003.04251}, Mar. 2020.

\bibitem{iuse_27} W. Dettling, ``How to teach blockchain in a business school,'' \textit{Springer Business information Systems and Technology, 2018}.


\bibitem{iuse_28} C. Dietrich, ``Wireless communication testbeds for authentic STEM learning,'' \textit{National Science Foundation Grant, \#1432416}, Aug. 2014. [Online]. Available: \url{https://nsf.gov/awardsearch/showAward?AWD_ID=1432416}

\bibitem{iuse_45} J. A. Sheridan, S. Kim, R. Goff, V. Marojevic, N. Polys, A. Mohammed, and C. Dietrich, ``Instructional strategies and design for immersive wireless communication tutorials and exercises,'' in \textit{Proc. ASEE Annual Conference 2017}.

\bibitem{iuse_29} Website of The Game Studio. [Online]. Available: \url{https://www.american.edu/gamelab/studio.cfm}

\bibitem{iuse_30} Website of The City University of New York (CUNY) Games Network. [Online]. Available: \url{https://games.commons.gc.cuny.edu/teaching-with-games/}

\bibitem{iuse_31} Website of Game-Based Education \& Advanced Research Studies (GEARS) Labs. [Online]. Available: \url{http://daytonabeach.erau.edu/about/labs/game-based-education-and-advanced-research}

\bibitem{iuse_32} Website of The A-Games Project. [Online]. Available: \url{http://gamesandlearning.umich.edu/a-games/key-findings/survey-report/digital-game-use/}

\bibitem{iuse_33} M. Spiegelman and R. Glass, ``Gaming and learning: Winning information literacy collaboration,'' \textit{College \& Research Libraries News}, vol. 69, no. 9, Oct. 2008.

\bibitem{iuse_34} B. Kim, ``Harnessing the power of game dynamics: Why, how to, and how not to gamify the library experience,'' \textit{Collage \& Research Libraries}, vol. 73, no. 8, 2012.

\bibitem{iuse_35} P. S. Buffum, M. Frankosky, K. E. Boyer, E. Wiebe, B. Mott, and J. Lester, ``Leveraging collaboration to improve gender equity in a game-based learning environment for middle school computer science,'' in \textit{Proc. IEEE Research in Equity and Sustained Participation in Engineering, Computing, and Technology (RESPECT) 2015}.


\bibitem{iuse_55} S. Kim, ``Impacts of mobility on performance of blockchain in VANET,'' \textit{IEEE Access}, vol. 7, 2019.

\bibitem{iuse_56} B. Gros, ``Digital games in education: the design of games-based learning environments,'' \textit{Journal of Research on Technology in Education}, vol. 40, no. 1, 2007.

\bibitem{iuse_54} 3GPP, ``Study on channel model for frequencies from 0.5 to 100 GHz,'' \textit{3rd Generation Partnership Project (3GPP), Technical Report TR 38.901, v16.1.0}, Release 16, Jan. 2020. [Online]. Available: \url{https://portal.3gpp.org/desktopmodules/Specifications/SpecificationDetails.aspx?specificationId=3173}

\bibitem{verboom} J. Verboom and S. Kim, ``Stochastic analysis on downlink performance of coexistence between WiGig and NR-U in 60 GHz band,'' \textit{arXiv preprint arXiv:2003.01570}, Mar. 2020.

\bibitem{kabir} M. Kabir and S. Kim, ``5G or Wi-Fi for HA/DR in the 60 GHz Band?,'' \textit{arXiv preprint arXiv:2003.00499}, Mar. 2020.

\bibitem{iuse_57} G.-T. Nguyen and K. Kim, ``A survey about consensus algorithms used in blockchain,'' \textit{J. Inf. Process. Syst.}, vol. 14, no. 1, pp. 1–28, Feb. 2018.

\bibitem{iuse_58} S. Nakamoto, Bitcoin: A Peer-to-Peer Electronic Cash System, Oct. 2008. [Online]. Available: \url{https://bitcoin.org/bitcoin.pdf}

\bibitem{iuse_59} J. A. Kroll, I. C. Davey, and E. W. Felten, ``The economics of bitcoin mining, or bitcoin in the presence of adversaries,'' in \textit{Proc. WEIS}, Jun. 2013.

\bibitem{access19} S. Kim, ``Impacts of mobility on performance of blockchain in VANET,'' \textit{IEEE Access}, vol. 7, May 2019.


\bibitem{iuse_50} Website of Criteria for Accrediting Engineering Programs, 2020 – 2021. [Online]. Available: \url{https://www.abet.org/accreditation/accreditation-criteria/criteria-for-accrediting-engineering-programs-2020-2021/}

\bibitem{iuse_61} R. K. Raj and A. Parrish, ``Toward standards in undergraduate cybersecurity education in 2018,'' \textit{IEEE Computer}, Feb. 2018.


\end{thebibliography}
\end{document}